\documentclass[runningheads]{llncs}
\usepackage{bibnames}
\usepackage{caption}
\usepackage{algpseudocode}
\usepackage{amsmath}
\usepackage{amsfonts}
\usepackage[ruled,vlined,resetcount,linesnumbered]{algorithm2e}
\makeatletter
\newcommand{\nosemic}{\renewcommand{\@endalgocfline}{\relax}}% Drop semi-colon ;
\newcommand{\dosemic}{\renewcommand{\@endalgocfline}{\algocf@endline}}% Reinstate semi-colon ;
\makeatother

\usepackage{graphicx}
\usepackage{subcaption}
\begin{document}
\title{Minimum Membership Geometric Set Cover in the Continuous Setting}
\author{Sathish Govindarajan \and
Mayuresh Patle \and
Siddhartha Sarkar
}
\authorrunning{S. Govindarajan, M. Patle, and S. Sarkar}
\institute{Indian Institute of Science, Bengaluru, India}
% \email{\{gsat, mayureshpatle, siddharthas1\}@iisc.ac.in}}
%
\maketitle              % typeset the header of the contribution
\begin{abstract}
We study the minimum membership geometric set cover, i.e., MMGSC problem [SoCG, 2023] in the continuous setting.
In this problem, the input consists of a set $P$ of $n$ points in $\mathbb{R}^{2}$, and a geometric object $t$, the goal is to find a set $\mathcal{S}$ of translated copies of the geometric object $t$ that covers all the points in $P$ while minimizing $\mathsf{memb}(P, \mathcal{S})$, where $\mathsf{memb}(P, \mathcal{S})=\max_{p\in P}|\{s\in \mathcal{S}: p\in s\}|$. 

For unit squares, we present a simple $O(n\log n)$ time algorithm that outputs a $1$-membership cover. We show that the size of our solution is at most twice that of an optimal solution. We establish the NP-hardness on the problem of computing the minimum number of non-overlapping unit squares required to cover a given set of points. This algorithm also generalizes to fixed-sized hyperboxes in $d$-dimensional space, where an $1$-membership cover with size at most $2^{d-1}$ times the size of a minimum-sized $1$-membership cover is computed in $O(dn\log n)$ time. Additionally, we characterize a class of objects for which a $1$-membership cover always exists. For unit disks, we prove that a $2$-membership cover exists for any point set, and the size of the cover is at most $7$ times that of the optimal cover. 
For arbitrary convex polygons with $m$ vertices, we present an algorithm that outputs a $4$-membership cover in $O(n\log n + nm)$ time. 
% For equilateral triangles, we demonstrate that a membership of $3$ is sufficient.
\end{abstract}
\section{Introduction}
\label{sec:intro}
The minimum membership geometric set cover problem has attracted significant interest in computational geometry due to its relevance in applications such as wireless networks, where minimizing interference is crucial \cite{socg23_minply,Biedl_Minply,Durocher_Mondal_Minply,erlebach_soda_minply,10.1007/978-3-031-52213-0_8}. Traditionally, much of the research on set cover problems has focused on discrete settings, where both the points to be covered and the covering objects are confined to predefined positions. However, many real-world scenarios require continuous flexibility in the placement of covering objects, leading to the study of the continuous variant of the problem.

The continuous geometric set cover problem for unit disks is a well-studied, classical problem in computational geometry. Its objective is to cover a given set of points with the smallest number of unit disks. In particular, a well-known PTAS exists for this problem based on the Hochbaum-Maass shifting strategy \cite{Hochbaum_Maass_covering}. Furthermore, there are numerous practical, fast constant-factor approximation algorithms \cite{BINIAZ20178,Bronnimann1995463,Fu2007317,GONZALEZ1991181}.

\begin{definition}(Ply)\label{def:ply}
    Given a set $\mathcal{S}$ of geometric objects, the ply of $\mathcal{S}$, denoted by $\mathsf{ply}(\mathcal{S})$, is $\max_{p\in \mathbb{R}^2}|\{s\in \mathcal{S}: p\in s\}|$.
\end{definition}

\begin{definition}(Membership)\label{def:memb}
    Given a set $P$ of points and a set $\mathcal{S}$ of geometric objects, the membership of $P$ with respect to $\mathcal{S}$,  denoted by $\mathsf{memb}(P, \mathcal{S})$, is $\max_{p\in P}|\{s\in \mathcal{S}: p\in s\}|$.
\end{definition}

\subsection{Our Contribution}
The Minimum-Membership Geometric Set Cover (MMGSC) problem is well studied in the discrete setting where both points and objects are given as input. In this paper, we initiate the study of the minimum membership geometric set cover (MMGSC) problem in the continuous setting. 
In this problem, the input consists of a set $P$ of $n$ points in $\mathbb{R}^{2}$, and a geometric object $t$, the goal is to find a set $\mathcal{S}$ of translated copies of $t$ that covers all the points in $P$, while minimizing $\mathsf{memb}(P, \mathcal{S})$, where $\mathsf{memb}(P, \mathcal{S})=\max_{p\in P}|\{s\in \mathcal{S}: p\in s\}|$. We present the following results on this problem. 
\begin{enumerate}
    \item $1$-membership Hypercube Cover: For unit intervals in one dimension, we give an exact algorithm that constructs a $1$-membership cover in $O(n\log n)$ time. Using this algorithm, we construct a $1$-membership cover for unit squares, and show that the size of the cover is a $2$-approximation to the optimum size. This algorithm also generalizes to (translates of) hyperboxes in $d$-dimension, where a $1$-membership cover with size at most $2^{d-1}$ times the size of a minimum-sized $1$-membership cover is computed in $O(dn\log n)$ time.
    \item We show that the problem of computing the minimum-size $1$-membership unit square cover is NP-hard by a reduction from $\mathsf{PLANAR3SAT}$.
    \item We show that a $1$-membership cover can be constructed if the geometric object $t$ tiles the plane. Moreover, for objects that do not tile the plane we show a point set for which a $1$-membership cover does not exist.
    \item For unit disks, we construct a $2$-membership cover, and show that the size of the cover is a $7$-approximation to the optimum.
    \item For convex polygons, we leverage homothetic approximations to achieve a $4$-membership cover.
\end{enumerate}
 
% We also report a $3$-membership cover that approximately minimizes the number of unit disks used. 
In this paper, we prove the bounds on ply, which implies the same bounds for membership. For example, in Section \ref{sec:hypercube}, we construct a $1$-ply hypercube cover. Since a $1$-ply cover is a set of non-overlapping objects, the membership of any point is at most $1$. Hence, this cover is a $1$-membership cover.

\subsection{Related Work}
\label{subsec:related_work}
Minimum Membership Geometric Set Cover (MMGSC) problem in the discrete setting (both points and objects are given as input) was introduced by Erlebach et al \cite{erlebach_soda_minply}, who presented NP-hardness and approximation results. A related problem is Minimum Ply Geometric Set Cover (MPGSC), introduced by Biedl et al \cite{Biedl_Minply}. They prove that the problem is NP-hard for unit squares and unit disks. Also, they gave a polynomial-time $2$-approximation when the minimum ply for the instance is a constant. Durocher et al. presented the first constant approximation algorithm for the MPGSC problem with unit squares \cite{Durocher_Mondal_Minply}.
Bandyapadhyay et al. introduced the \textit{generalized} MMGSC problem, which is a generalization of both MMGSC and MPGSC. They gave a polynomial-time $144$-approximation algorithm for unit squares \cite{socg23_minply}. 
Govindarajan and Sarkar later improved the approximation factor to $16$ \cite{10.1007/978-3-031-52213-0_8}. The Unique Coverage problem is another related problem, which was introduced by Demaine et al. for the set systems \cite{demaine2008combination}. Erlebach and van Leeuwen gave the first set of results for geometric unique coverage with unit squares and unit disks \cite{erlebach_soda_minply}. They showed that the unique coverage of unit disks is NP-hard and presented an $18$-approximation algorithm with $O(n^{3}m^{8})$ runtime. For unit squares, they gave a $4$-approximation algorithm. Later, van Leeuwen established NP-hardness of the unique coverage of unit squares as well, and gave a $2$-approximation algorithm for the problem \cite{van2009optimization}.

\section{$1$-ply Hypercube Cover}
\label{sec:hypercube}
In this section, we construct a $1$-ply cover for hypercubes in $d$ dimension using the $1$-ply cover in $(d-1)$ dimension. 

\subsection{Unit Interval Cover}
\label{sec:interval}
First, we consider the setting in one dimension where $P$ is a set of points on the $x$-axis. We show that a set $S$ of disjoint unit-length intervals that cover all points in $P$ can always be found. 

\begin{lemma}\label{lemma:interval-cover}
Given a set $P$ of $n$ points on the $x$-axis, a $1$-ply cover with minimum number of unit intervals can be computed in $O(n\log n)$ time.
\end{lemma}

\begin{proof}
    Any two distinct input points on the $x$-axis are separated by a non-zero distance. We construct a $1$-ply interval cover by sweeping the $x$-axis from left to right. Whenever an uncovered point $p\in P$ is encountered, add a unit interval $s$ to the solution set $\mathcal{S}$, where $p$ is the left endpoint of $s$. This algorithm, referred to as $\mathsf{Separate}(P)$, produces an ordered set $\mathcal{S}$ of non-overlapping unit intervals covering $P$, from left to right, in $O(n\log{n})$ time. The optimality of this greedy algorithm can be proved by a standard \textit{stay ahead} argument. 
    
    Let $OPT$ be a minimum cardinality $1$-ply unit interval cover for $P$. Let $s_i$ and $opt_{i}$ denote the $i^\text{th}$ interval in $S$ and $OPT$, respectively (according to the left-to-right order). For $i>=1$, let $S_{i}=\{s_1, \ldots, s_i\}$, and $OPT_{i}=\{opt_1, \ldots, opt_i\}$. We claim that $S_{i}$ covers all the points covered by $OPT_{i}$, for all $i\in |OPT|$. We prove this claim inductively. By construction, $s_1$ starts at the leftmost point of $P$, which must be covered by $opt_{1}$. For $i=1$, the claim holds since both $s_1$ and $opt_{1}$ are unit length. Let the claim be true for $S_{i-1}$. Let $p$ be the leftmost point not covered by $S_{i-1}$. Assume for a contradiction that $S_{i}$ does not cover all points covered by $OPT_{i}$. Thus, by the inductive hypothesis, $opt_i$ must cover all points of $s_i$ plus at least one additional point to the right of $s_i$. By construction, $s_i$ starts at $p$. Given that $opt_i$ contains $p$ and ends strictly after $s_i$ ends, this scenario is impossible because the intervals are of unit length. Hence, a contradiction. Therefore, $S$ is a minimum-sized $1$-ply unit interval cover. \qed
\end{proof}

\subsection{Unit Square Cover}\label{sec:square-cover}
Given a set $P$ of points in the plane, the goal is to produce a set $S$ of axis-aligned unit squares that cover all the points in $P$ while minimizing the ply. Unless stated otherwise, a square refers to a unit square.

\begin{theorem} \label{lemma:square-cover}
Given a set $P$ of $n$ points in the plane, a $1$-ply unit square cover can be computed in $O(n\log{n})$ time.
\end{theorem}

\begin{proof}
To generate a $1$-ply unit square cover for a given set $P$ of $n$ points in the plane, apply the $\mathsf{Separate}$ algorithm (defined in the proof of Lemma \ref{lemma:interval-cover})  on the $x$-coordinates of the points, which distributes the points into non-overlapping vertical strips; see Fig. \ref{fig:square-continuous}(a). Again, apply the $\mathsf{Separate}$ algorithm with unit intervals on the $y$-coordinates of the points within each strip to split it into squares; see Fig. \ref{fig:square-continuous}(b). 

This algorithm is referred to as $\mathsf{SquareCover}$. Suppose, $k$ vertical strips are generated and the $i$-th vertical strip contains $n_i$ points such that $\sum_{i\in [k]}n_i = n$. Then the running time of $\mathsf{SquareCover}$, ignoring multiplicative constants, is $n\log n+\sum_{i\in [k]} n_i \log n_i \leq n\log n + \log n \sum_{i\in [k]} n_i = 2\cdot n\log n$. \qed
\begin{figure}[ht!]
    \centering
    \includegraphics[width=8cm]{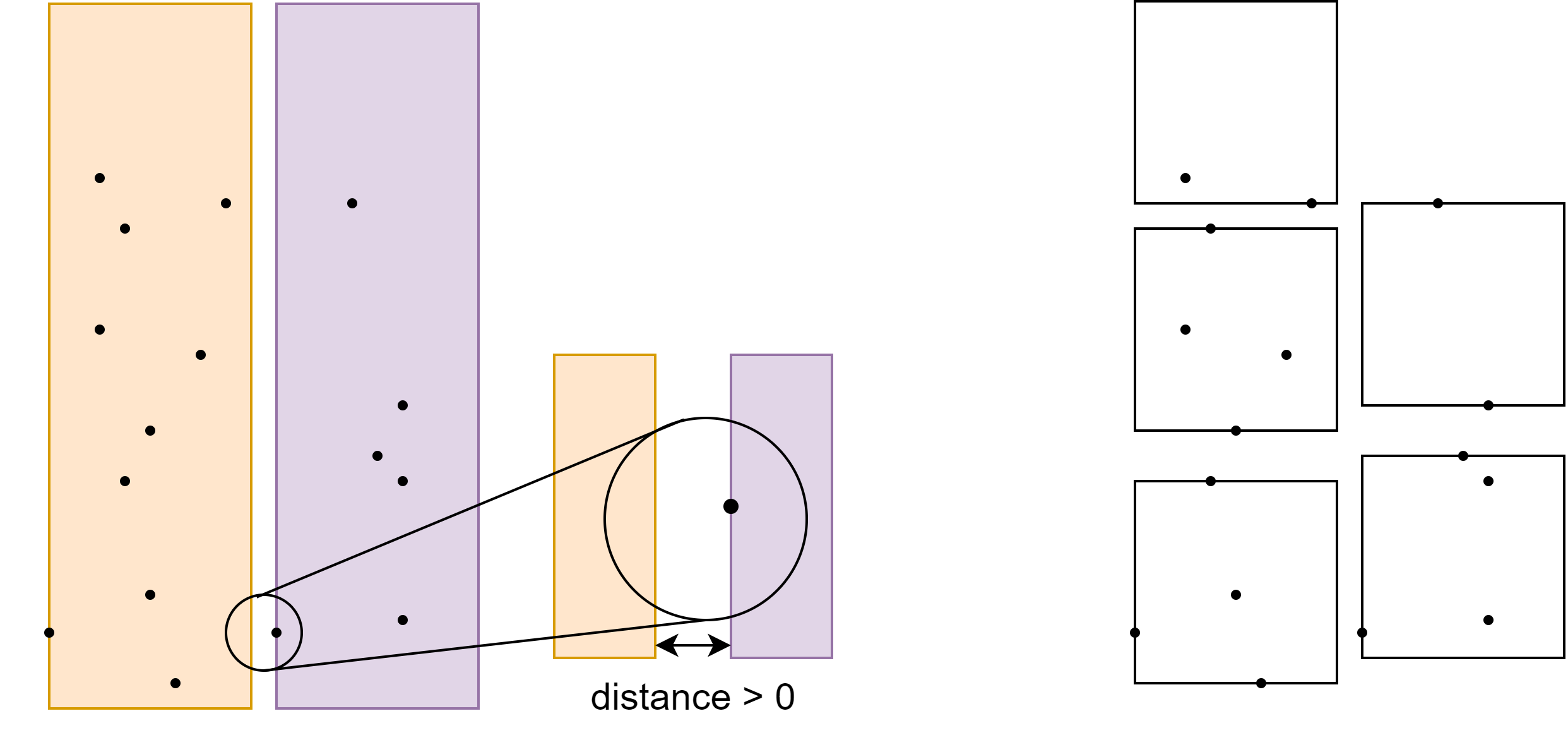}
    \caption{(a) Partitioning into vertical strips. \quad \quad (b) $1$-ply unit square cover.}
    \label{fig:square-continuous}
\end{figure}
\end{proof}

\begin{theorem} \label{lemma:2-approx-square}
The size of the cover generated by the $\mathsf{SquareCover}$ algorithm is at most twice the size of any minimum-sized $1$-ply unit square cover for a given set $P$ of points in the plane.
\end{theorem}

\begin{proof}
The $\mathsf{Separate}$ algorithm divides the points into vertical strips of width $1$. These strips are enumerated from left to right as $V_1, V_2, V_3, \ldots$. Let $O$ be the union of the odd-indexed strips, and $E$ be the union of the even-indexed strips. Since there is a gap between $V_i$ and $V_{i+1}$, as well as between $V_{i+1}$ and $V_{i+2}$, any unit square that covers a point in $V_{i}$ cannot cover a point in $V_{i+2}$. Therefore, if only the points in $O$ (resp. $E$) were given, an optimal cover could be found by finding the optimal covers for each strip in $O$ (resp. $E$) individually. For a vertical strip, the problem reduces to the $1$-ply unit interval cover. So applying the $\mathsf{Separate}$ algorithm on the $y$-coordinates of points in a strip gives an optimal cover for a vertical strip. Thus, we have an optimal cover for the points in $O$ (resp. $E$). Let $S_O$ and $S_E$ be the sets of squares obtained by $\mathsf{SquareCover}$ for the points in $O$ and $E$, respectively. Let $OPT$ be a minimum-sized $1$-ply unit square cover for $P$. Since the set of points contained in $O$ (resp. $E$) is a subset of $P$, therefore, $|S_O| \leq |OPT|$, and $|S_E| \leq |OPT|$. Thus $|S_O| + |S_E| \leq 2|OPT|$. Therefore, the set cover $S_O \cup S_E$ is at most twice as large as $OPT$. \qed
\end{proof}

\noindent \textit{Remark.} These theorems can be generalized to apply to axis-aligned rectangle cover, with fixed-sized rectangles, say with dimensions $a \times b$, by separating on $x$-direction with $a$-length intervals and separating on $y$-direction with $b$-length intervals.

\subsection{Hypercube Cover}
\label{sec:hypercube-cover}

Given a set $P$ of points in the $d$-dimensional space, the goal is to produce a set $S$ of axis-aligned unit hypercubes that cover all points in $P$ while minimizing the ply.

\begin{theorem}\label{lemma:hypercuber-cover}
    Given a set $P$ of $n$ points in $d$-dimensional space, a $1$-ply unit hypercube cover, which is at most $2^{d-1}$ times the size of any minimum-sized $1$-ply cover, can be generated in $O(d \cdot n\log{n})$ time.
\end{theorem}

This theorem can be derived by generalizing the results of the minimum ply unit square cover problem inductively. We prove two lemmas from which Theorem \ref{lemma:hypercuber-cover} follows directly. First, we define some terms.
\begin{definition}[$d$-dimensional wall]
    For $x\in \mathbb{R}$, an orthogonal range of the form $(-\infty, \infty)^{d-1} \times [x, x+1]$, is called a $d$-dimensional wall.
\end{definition}

\begin{definition}[$d$-dimensional projection of a point]
    A $d$-dimensional point, obtained by ignoring the coordinates in higher dimensions (if any) while preserving the coordinates in the first $d$ dimensions.
\end{definition}

\begin{lemma}\label{claim:hypercube-cover}
    Given a set $P$ of $n$ points in $d$-dimensional space, a $1$-ply unit hypercube cover can be generated in $O(d \cdot n\log{n})$ time. 
\end{lemma}

\begin{proof}
Consider the $\mathsf{HypercubeCover}$ algorithm that takes the point set $P$ and the number of dimensions $d$ as input and returns a set $\mathcal{S}$ of axis-aligned $d$-dimensional unit hypercubes that cover the $d$-dimensional projection of $P$.

\begin{algorithm}[ht!]
    \caption{$\mathsf{HypercubeCover}(P, d)$}
    \label{algo:hypercube-cover}
    \SetKwComment{Comment}{$\triangleright$ }{ }
    \If{$d = 2$}{Return $\mathsf{SquareCover(P)}$}
    $ \mathcal{S} \gets \emptyset$ \Comment*{Set of hypercubes in the cover}
    Sort $P$ in non-decreasing order based on the $d^\text{th}$ coordinate (for efficient identification of points in the same wall)\;
    $P_d \gets$ Set of $d^\text{th}$ coordinates of points in $P$\;
    $I_d \gets \mathsf{Separate}(P_d)$\;
    \For{$drange \in I_d$}{
        $P' \gets$ All points in $P$ having their $d^\text{th}$ coordinate in $drange$\;
        $S_{d-1}\gets HypercubeCover(P', d-1)$ \;
        \For{$box \in S_{d-1}$}{
            Insert $box \times drange$ into $\mathcal{S}$\;
        }
    }
    Return $\mathcal{S}$ \;
\end{algorithm}
The base case for $d=2$ generates a $1$-ply unit square cover. For $d > 2$, we assume that $\mathsf{HypercubeCover}(P, d-1)$ generates a $1$-ply $(d-1)$-dimensional unit hypercube cover. We observe that $I_d$, which is a $1$-ply unit interval cover for the $d^\text{th}$ dimension, assigns each point to its respective wall. For a wall corresponding to $drange \in I_d$, containing subset $P' \subseteq P$ of points, we have a $1$-ply cover obtained using $\mathsf{HypercubeCover}(P', d-1)$, which is $S_{d-1}$. Thus, we conclude that $\{box \times drange\ |\ \forall box \in S_{d-1} \}$ will be a $1$-ply unit hypercube cover for the $d$-dimensional wall. No two walls intersect or touch since $I_d$ is a $1$-ply cover. Hence, the combined solution of all walls is a $1$-ply hypercube cover for $P$. 

Let $T_d(|P|)$ denote the time complexity of $\mathsf{HypercubeCover}(P, d)$. By Theorem \ref{lemma:square-cover}, we have the base case, $T_{2}(n) = 2\cdot n\log n$. Suppose, there are $k$ $d$-dimensional walls and the $i$-th one contains $n_i$ points such that $\sum_{i\in [k]}n_i = n$. Each recursive invocation of the $\mathsf{HypercubeCover}$ algorithm includes a sorting operation, leading to the subsequent recurrence.
\begin{equation}
T_d(n) = n \log n + \sum_{i \in [k]} T_{d-1}(n_i) \\
= d \cdot n \log n
\end{equation}

The second equality follows from the base case of $d=2$. Thus, the overall time complexity of $\mathsf{HypercubeCover}(P, d)$ is $O(d\cdot n\log n)$, where $|P|=n$. \qed
\end{proof}

\begin{lemma} \label{claim:approx-hypercube}
    The cover generated by the $\mathsf{HypercubeCover}$ algorithm is at most $2^{d-1}$ times the size of any minimum-sized $1$-ply unit hypercube cover for a given set $P$ of points in $d$-dimensional space.
\end{lemma}

\begin{proof}
    We prove this claim by induction on the number of dimensions $d$. The claim is true for $d=2$, as proved in Theorem \ref{lemma:2-approx-square}. Assume that the claim is true for $d=i-1$, and let us prove it for $d=i$. 
    
    The algorithm for the unit hypercube cover in $d$-dimensional space distributes the points into several disjoint walls. Let $W_1, W_2, W_3, \ldots$ denote the walls enumerated in increasing order of their ranges in $d^\text{th}$ dimension, and let $O$ be the union of odd-indexed walls and $E$ be the union of even-indexed walls. 
    
    A wall has $d^\text{th}$-dimension range with unit length. Thus, for points in a single wall, the size of the smallest $d$-dimensional $1$-ply hypercube cover is the same as the size of the smallest $(d-1)$-dimensional $1$-ply hypercube cover for $(d-1)$-dimensional projections of those points. Thus, by construction and inductive hypothesis, we have a $2^{d-2}$ approximation for the points in each of the $d$-dimensional walls. Now, by the separation of walls due to gaps in $S_d$, and by the presence of a wall $W_{i+1}$ between $W_i$ and $W_{i+2}$, any hypercube cannot cover two points such that one lies in $W_i$ and the other lies in $W_{i+2}$. Hence, we have a $2^{d-2}$ approximation for the points in $O$ and $E$, respectively. 
    
    Let $OPT_O$ and $OPT_E$ be the optimal solutions for points in $O$ and $E$, respectively, and let $OPT$ be the optimal solution for $P$. Since $P$ contains points in $O \cup E$, we have $|OPT_O| \leq |OPT|$ and $|OPT_E| \leq |OPT|$. 
    
    Let $S_O$ and $S_E$ be the sets of hypercubes generated by the $\mathsf{HypercubeCover}$ algorithm for the regions corresponding to $O$ and $E$, respectively. Thus,
    \begin{equation*}
    \begin{split}
        & |S_O| \leq 2^{d-2} |OPT_O| \leq 2^{d-2} |OPT|, 
        |S_E| \leq 2^{d-2} |OPT_E| \leq 2^{d-2} |OPT| \\
        \implies & |S_O \cup S_E| = |S_O|+|S_E| \leq 2^{d-1}|OPT|.
    \end{split}
    \end{equation*}
    Therefore, the hypercube cover generated by the $\mathsf{HypercubeCover}$ algorithm is at most $2^{d-1}$ times the size of any minimum-sized $1$-ply unit hypercube cover. \qed
\end{proof}

\noindent \textit{Remark.} This theorem can be further extended to d-dimensional axis-aligned hyperbox cover, with dimensions $l_1 \times l_2 \times \cdots \times l_d$, by determining the $l_i$-length interval cover while separating along the $i^\text{th}$ dimension.

\subsection{NP-hardness of Minimum Size $1$-Ply Unit Square Cover}
\label{sec:hardness}
Let us first define the minimum size $1$-ply unit square cover problem formally. 

\begin{figure}[ht!]
    \centering
    \includegraphics[width=7cm]{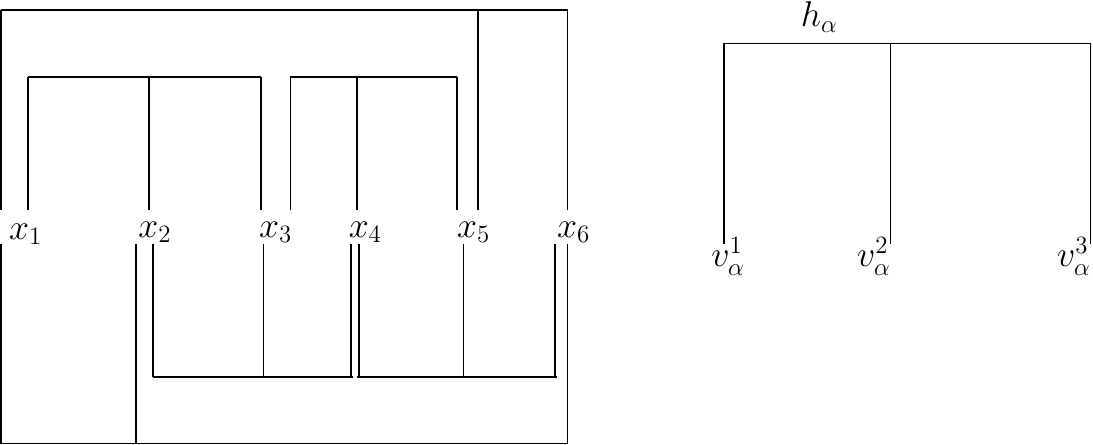}
    \caption{The left figure shows a planar embedding of a $\mathsf{PLANAR3SAT}$ instance. The right figure shows a comb structure for a clause $\alpha$.}
    \label{fig:rect_embedding}
\end{figure}

\begin{definition}[Minimum Size $1$-Ply Set Cover of Unit Squares]\label{def:minsize_1ply_us}
Given a set of $n$ points $P$ on $\mathbb{R}^2$, the goal in $\mathsf{MS1P}$-$\mathsf{SC}$-$\mathsf{US}$ is to cover $P$ with the minimum number of non-overlapping unit squares.
\end{definition}
We prove that the above problem is NP-hard via a reduction from $\mathsf{PLANAR3SAT}$, which is known to be NP-hard \cite{planar3sat_lichtenstein_1982}. Recall that $\mathsf{3SAT}$ asks whether there exists a truth assignment to the variables of a given $\mathsf{3SAT}$ formula $\varphi$ that satisfies it. $\mathsf{PLANAR3SAT}$ adds a geometric constraint that the variable-clause incidence graph of $\varphi$ must be planar. There are many ways to embed the incidence graph of a $\mathsf{PLANAR3SAT}$ formula on the plane without edge crossings. Knuth and Raghunathan show how the graph can be laid out (in polynomial time) such that variables correspond to points on the $x$-axis and clauses correspond to non-crossing three-legged ``combs'' above or below the $x$-axis \cite{knuth_raghunathan_1992_planar3sat}. First, place all the variable nodes along a horizontal line, in order. Then, for each clause, connect its three variable nodes with rectilinear (i.e., right-angled) non-crossing line segments. Visually, for each clause, the rectilinear connections form a ``three-legged comb''. Refer to Fig. \ref{fig:rect_embedding}. See \cite{planar_3sat_embedding_GD2008} for a relevant reduction. 

\begin{theorem}
        $\mathsf{MS1P}$-$\mathsf{SC}$-$\mathsf{US}$ is NP-hard.
\end{theorem}
\begin{proof}
Given a $\mathsf{PLANAR3SAT}$ instance $\varphi$ with $n$ variables and $m$ clauses, we construct in polynomial time an instance $P_{\varphi}$ of $\mathsf{MS1P}$-$\mathsf{SC}$-$\mathsf{US}$ such that the following holds: 
$P_{\varphi}$ can be covered by at most $k$ non-overlapping unit squares if and only if $\varphi$ is satisfiable. We fix $k$ later. First, we place some \textit{guiding} unit squares on the plane that, in turn, decide the points in $P_{\varphi}$. These squares will be colored red or blue. The squares of the same color will be pairwise disjoint. All but $m$ points in $P_{\varphi}$ will be placed in the intersection regions of the overlapping unit squares. The placement of the \textit{guiding} unit squares and the points in $P_{\varphi}$ is done in steps, leading to some gadgets, as described below.

    \textbf{Variable gadget}. A variable gadget is a horizontal chain of an even number of unit squares. 
    There could be two types of unit squares in a variable gadget, namely, `variable squares' and `separating squares'. Let $x_i$ be a variable in $\varphi$ that appears in $k_i$ clauses. We put $k_i + 2(k_i-1)=3k_i-2$ pairs of unit squares sequentially, forming a horizontal chain, where every two consecutive squares intersect. Essentially, between every two `variable square' pairs, we have a quadruple of `separating squares'. To be precise, the chain starts with a pair of `variable squares'. These are immediately followed by a quadruple of `separating squares'. The pattern continues as alternating between a pair of `variable squares' and a quadruple of `separating squares', until we place $k_i$ pairs of `variable squares'. The separating squares ensure enough spacing among the vertical chains of squares (to be placed later). See Fig. \ref{fig:clause_gadget}(a). 
    
    The unit squares in the chain alternate colors, with the leftmost being red. 
    Two points are placed within each rectangular intersection region of two consecutive squares (i.e., a red and a blue square): one at the bottom-left and the other at the top-right corner.
    These points are termed the \textit{variable points}. Furthermore, certain variable squares are moved vertically to appropriately connect with the \textit{clause squares}.

\begin{figure}[ht!]
  \centering
  % Top row: fig1 and fig2 side by side
  \begin{subfigure}[t]{0.45\textwidth}
    \centering
    \includegraphics[width=\linewidth]{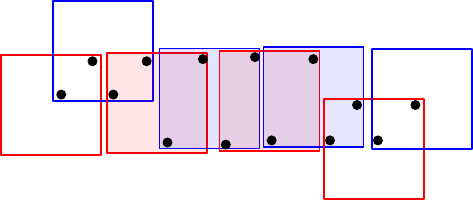}
    \caption{A variable gadget for a variable that appears in two clauses. The \textit{separating} squares are shaded.}
    \label{fig:fig1}
  \end{subfigure}%
  \hfill
  \begin{subfigure}[t]{0.45\textwidth}
    \centering
    \includegraphics[width=\linewidth]{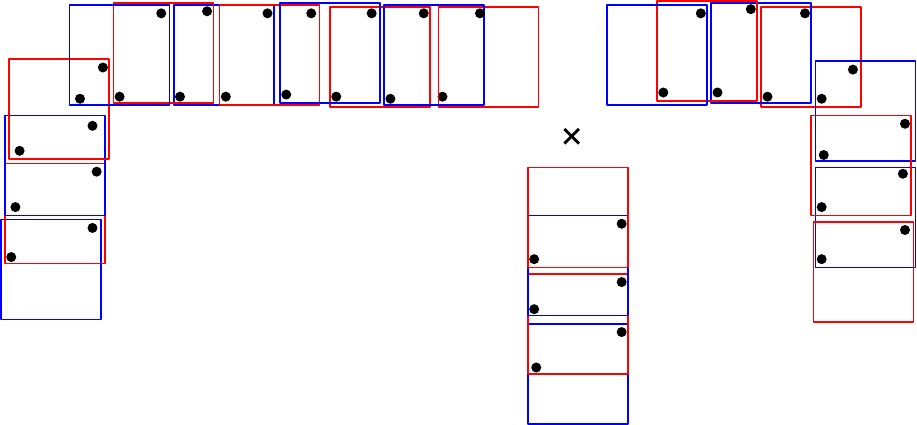}
    \caption{A clause gadget where the central clause point is denoted by a cross.}
    \label{fig:fig2}
  \end{subfigure}

  \vspace{1em}

  % Bottom row: fig3 full width
  \begin{subfigure}[t]{0.8\textwidth}
    \centering
    \includegraphics[width=\linewidth]{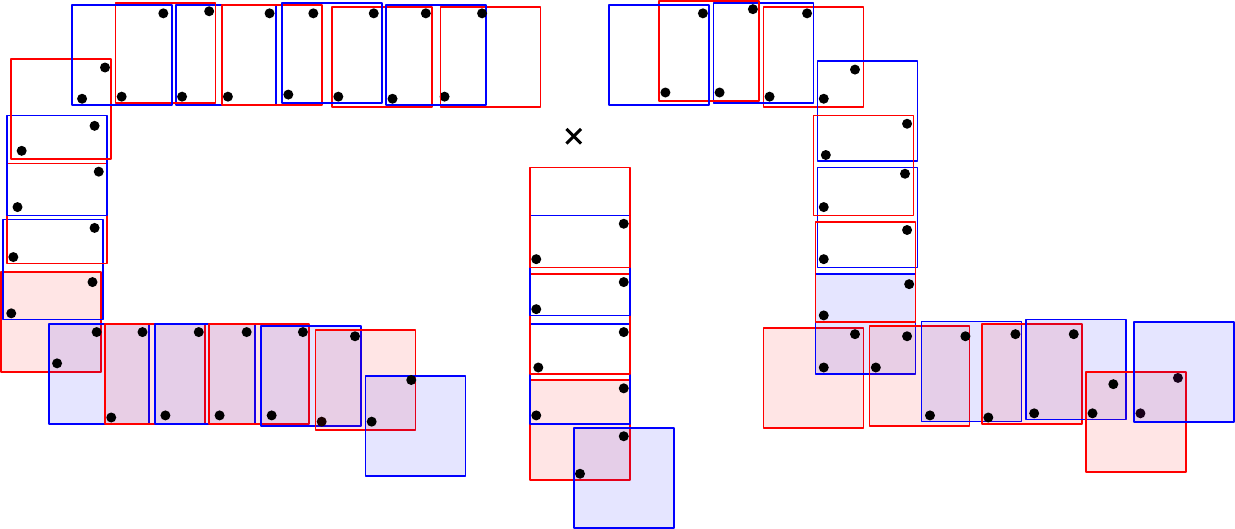}
    \label{fig:fig3}
    \caption{Connecting a clause gadget with the corresponding variable gadgets. The shaded squares correspond to the variable gadgets.}
  \end{subfigure}

  \caption{Depicting a clause $(\neg x_i \lor \neg x_j \lor x_k)$. The variable gadgets for $x_i, x_j, x_k$ are drawn from left to right; $x_i$ appears in its positive form in some other clause; $x_j$ appears only in $\alpha$; and $x_k$ appears in its negative form in some other clause.}
  \label{fig:clause_gadget}
\end{figure}

    \textbf{Clause gadget}. For every clause $\alpha$ in $\varphi$, there is a $3$-legged comb, i.e. a horizontal segment $h_{\alpha}$ and three vertical segments $v^1_{\alpha}, v^2_{\alpha}, v^3_{\alpha}$, in the planar embedding (see Fig. \ref{fig:rect_embedding}). 
    We place squares along $h_{\alpha}, v^{1}_{\alpha}, v^{2}_{\alpha},$ and $v^{3}_{\alpha}$ and color them red and blue alternately. 
    Every two adjacent squares intersect.
    Along $h_{\alpha}$, we place two subchains of squares with a gap in the junction of $h_{\alpha}$ and $v^{2}_{\alpha}$ (i.e., the middle leg of the comb).
    Thus, a clause gadget has two horizontal chains and three vertical chains of guiding unit squares. These squares are called \textit{clause squares}. See Fig. \ref{fig:clause_gadget}(b).
    As in the variable gadget, two points are placed within each rectangular intersection region of two intersecting squares. 
    These points are termed as \textit{clause points}.
    We place a point $p_{\alpha}$ near the intersection of $h_{\alpha}$ with $v^2_{\alpha}$ in a specific way, to be made precise shortly. The point $p_{\alpha}$ is called the \textit{central clause point} of $\alpha$.  
    If a variable $x_i$ appears in its positive form in $\alpha$, then a blue square of the variable gadget of $x_i$ intersects a red square of the corresponding vertical chain of $\alpha$, as shown in Fig. \ref{fig:clause_gadget}(c). 
    If a variable appears in its negated form, then a red square of the variable gadget intersects a blue square of the corresponding vertical chain, as shown in Fig. \ref{fig:clause_gadget}(c). 
    At this point, we do not worry about the exact number of squares in a chain.  
    The \textit{central clause squares} are defined as the three squares nearest to $p_{\alpha}$, each located at the ends of the corresponding chains of clause squares for $\alpha$.
    We ensure that the positioning of the point $p_{\alpha}$ and the central clause squares respects the following properties.
    \begin{itemize}
        \item For each clause $\alpha$, any square (on the plane) containing $p_{\alpha}$ intersects at least one of the corresponding central clause squares.
        \item A square containing $p_{\alpha}$ can be drawn, intersecting only one central clause square and no other guiding squares.
    \end{itemize}
This completes the construction of the set $P_{\varphi}$ of points. It is easy to see that the reduction takes polynomial time.

We now prove that the $\mathsf{PLANAR3SAT}$ formula $\varphi$ is satisfiable if and only if there exists a set of at most $k$ non-overlapping unit squares covering $P_{\varphi}$. Here, $k=m+c$, where $m$ is the number of clauses in $\varphi$ and $c$ is half the number of guiding squares placed during the reduction.

First, consider the forward direction. Suppose that $\varphi$ is a satisfiable formula. 
We place $k$ non-overlapping unit squares that cover the point set $P_{\varphi}$ as follows. Consider a satisfying truth assignment $\delta_{\varphi}$ for $\varphi$. 
For each variable $x_i$ which is set to True according to $\delta_{\varphi}$, select the red squares in the corresponding variable gadget. 
For each of the remaining variables, select the blue squares in the corresponding variable gadget. 
This, in turn, determines the squares to be chosen from the clause chains. 
Since every clause $\alpha$ is satisfied, for each clause, at least one literal gets evaluated to True. 
Hence, the construction ensures that at least one central clause square of $\alpha$ is not selected, leaving enough room for placing a non-overlapping square that covers $p_{\alpha}$.
Thus, $k=c+m$ non-overlapping squares are selected to cover $P_{\varphi}$.

Now consider the reverse direction. Let the formula $\varphi$ be a no-instance, i.e., it is not satisfiable. 
Suppose for a contradiction that $\mathcal{S}$ is a $1$-ply unit square cover of $P_{\varphi}$ of size at most $k=m+c$. 
Define the {\em budget} for a variable or clause as half the number of corresponding guiding squares placed during the reduction.
By construction, for each variable $x_i$, only two distinct square patterns exist that can cover the \textit{variable points} of $x_i$, while not exceeding the budget for $x_i$.
The same is true for a clause $\alpha$ in $\varphi$.
Moreover, covering the central clause points in $P_{\varphi}$ requires at least $m$ additional unit squares.
Since $\varphi$ is not satisfiable, for any truth assignment, there exists an unsatisfied clause, say, $C_{i}$. 
Since $\mathcal{S}$ must respect the budget for each variable/clause in $C_{i}$, to cover the points in $C_{i}$ within the budget, it is necessary to choose all the three central clause squares of $C_{i}$.
Hence, there is at least one clause for which the number of unit squares required will exceed the budget. Thus, the number of non-overlapping unit squares required to cover $P_{\varphi}$ is strictly more than $k$. \qed
\end{proof}

\section{Minimum Ply Cover using Tiling Objects}\label{sec:tiling-objects}
In this section, we characterize the minimum ply cover of objects that tile the plane. An object is called a {\em tiling object} if the entire plane can be tiled using translated copies of the object. In other words, the plane is an interior-disjoint union of translated copies of the object. Examples of tiling objects are a square and a regular hexagon. We give the following characterization:

\begin{theorem}
Given a set $P$ of $n$ points in the plane and an object $t$, a $1$-ply cover of $P$ with translated copies of $t$ exists if and only if $t$ is a tiling object.
\end{theorem}

\begin{proof}
Let $t$ be a tiling object and consider a tiling of the plane with $t$. We obtain a $1$-ply cover by ensuring that the translated copies of $t$ are boundary disjoint as follows: Perform a uniform expansion of the boundary of $t$ by a small amount $\delta, \delta > 0$ to obtain an expanded object $t'$. Consider the tiling $T'$ of the plane using $t'$ such that the points of $P$ are at least $\delta$ distance away from the boundary of the tiling. Now, shrink the objects $t'$ in tiling $T'$ by an amount $\delta$ so that they become translated copies of $t$. As these translated copies are boundary-disjoint and they cover $P$, we get a $1$-ply cover of $P$.

To prove the \textit{only if} part, consider an object $t$ that is not a tiling object. Consider a packing of the plane using translated copies of $t$ that minimizes the size $s$ of the smallest hole. The size $s$ of a hole in a packing is defined as the diameter of the largest disk that can be inscribed within the hole. Let $P$ be a grid of points with grid cell size $< s$. $P$ cannot be covered using disjoint translated copies of $t$. \qed
\end{proof}

We now make an observation on the size of the $1$-ply cover of a tiling object $t$. Consider the $1$-ply cover $\mathcal{C}$ of $P$ consisting of non-empty objects of the tiling as constructed in the above proof. Let $m_t$ be the maximum number of objects in $\mathcal{C}$ that can be intersected by a translated copy of $t$.

\begin{lemma}\label{lemma:overlap_factor}
The size of the $1$-ply cover $\mathcal{C}$ is an $m_t$-approximation to the minimum-sized $1$-ply cover of $P$.
\end{lemma}
\begin{proof}
Each object in the minimum-sized $1$-ply cover $O$ of $P$ intersects at most $m_t$ objects in $\mathcal{C}$. Also, since each object in $\mathcal{C}$ is non-empty, it is intersected by some object in $O$. Thus, the size of the cover $\mathcal{C}$ is at most $m_t \vert O \vert$. \qed
\end{proof}

Remark: By the above lemma, we get a $4$-approximation for squares and regular hexagons.

\section{Unit Disk Cover}
\label{sec:disk}

Given a set $P$ of $n$ points in the plane, the objective is to produce a set $S$ of unit disks (disks with diameter $1$) that cover all points in $P$ while minimizing ply of $P$. In this section, we prove that a ply of $2$ is necessary and sufficient.

\subsection{$2$-Ply Unit Disk Cover}

\begin{lemma} \label{lemma:disk-cover}
    Given a set $P$ of $n$ points in the plane, a $2$-ply unit disk cover can be constructed, which is at most $7$ times the size of a minimum-sized $2$-ply cover.
\end{lemma}
\begin{proof}
    Generating a $2$-ply unit disk cover for a point set $P$ consists of two steps. The first step is to apply the $\mathsf{Separate}(P)$ algorithm to distribute the points in $P$ into boundary-disjoint vertical strips. This is achieved by using an interval length of $\frac{1}{\sqrt{2}}$ for the $x$-coordinates of the points. Next, the $\mathsf{Separate}(P)$ algorithm is again applied to distribute the points in $P$ into boundary-disjoint horizontal strips based on their $y$-coordinates. This construction ensures that all points lie within the intersection of these vertical and horizontal strips, resulting in square regions of side length $1/\sqrt{2}$. Therefore, a $1$-ply square cover is obtained.

The next step is to draw circumcircles over these squares, resulting in a unit disk cover, say $\mathcal{D}$. 
    %The overall time complexity of this process is $O(n\log n)$, resulting from the application of the $Separate(P)$ algorithm and binary search operations for each point.
The vertical and horizontal alignment of the squares in the $1$-ply cover ensures that the drawn circumcircles form a grid (Figure \ref{fig:disk_2_ply}). In this grid of circles, only vertical or horizontal neighbors can cross, but diagonally adjacent disks do not intersect or touch. In Fig. \ref{fig:disk_2_ply}, $|AB|, |BC|> 1/\sqrt{2}$, hence $|AC| = \sqrt{|AB|^2 + |BC|^2} > 1$. This property guarantees that $\mathcal{D}$ is a $2$-ply unit disk cover for the point set $P$.

We claim that any unit disk can intersect at most $7$ unit disks of $\mathcal{D}$. Suppose not. Then, the $8$ disks intersecting a unit disk, say $d$, must be from the $9$ disks shown in Fig. \ref{fig:disk_2_ply}. Hence, at least two diagonally opposite disks must be among them, which is a contradiction. 
Thus, by Lemma \ref{lemma:overlap_factor}, we get a $7$-approximation to the minimum-sized $2$-ply unit disk cover of $P$. \qed

    \begin{figure}[ht!]
        \centering
        \includegraphics[width=6cm]{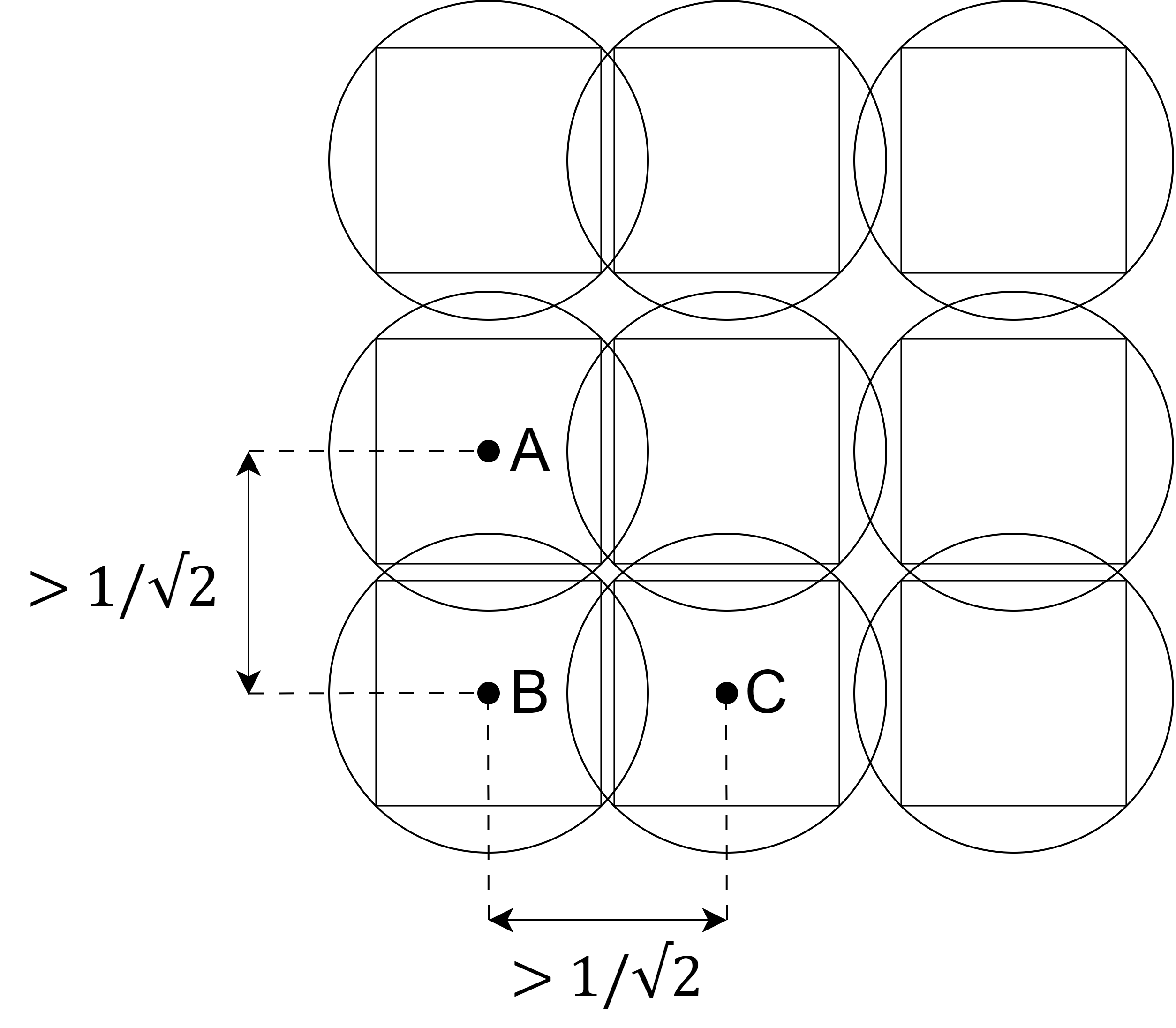}
        \caption{Illustration of a 2-ply unit disk cover configuration.}
        \label{fig:disk_2_ply}
    \end{figure}
\end{proof}

It is known that there is a set of points for which a $1$-ply unit disk cover cannot exist \cite{Aloupis2012CoveringPW}.

\section{Convex Polygon Cover}
\label{sec:polygon}

Let $P$ be a set of $n$ points in the plane and let $C$ be an arbitrary convex polygon with $m$ vertices given as a sorted array. The goal is to find a set $S$ of translations of $C$ to cover all points in $P$ while minimizing the ply.

We use the same terminology as in \cite{SCHWARZKOPF199877,lassak-homothety,all-shapes}. A pair of rectangles $(r, R)$ is called \textit{homothetic} if they are parallel and have the same aspect ratio (not necessarily axis-parallel). A homothetic pair $(r, R)$ is an \textit{approximating pair} for $C$ if $r \subseteq C \subseteq R$. That is, $r$ is enclosed in $C$, and $C$ is enclosed in $R$, see Fig. \ref{fig:convex}(a). Let $\lambda(r, R)$ be the smallest ratio of the length of $R$ to the length of $r$ over all convex shapes. It was shown in \cite{SCHWARZKOPF199877,lassak-homothety} that $\lambda(r, R) \leq 2$ for any convex shape. Schwarzkopf et al. \cite{SCHWARZKOPF199877} also showed that if $C$ is a convex polygon with $m$ vertices given as a sorted array, then an approximating pair of rectangles with sides at most twice as long as each other can be computed in time $O(\log^2m)$.

\begin{theorem} \label{theorem:convex-cover}
Given a set $P$ of $n$ points in the plane and an arbitrary convex polygon $C$ with $m$ vertices given as a sorted array, there exists an algorithm that can generate a set of translations of $C$ to cover all points in $P$ with a ply value of at most $4$ in $O(n\log{n} + mn)$ time.
\end{theorem}

\begin{proof}
We start by finding an approximating pair $(r, R)$ for $C$ where $\lambda(r, R) \leq 2$, and assume $\lambda(r, R) = 2$ for simplicity (this can be achieved by shrinking $r$). This step takes $O(\log^2m)$ time. Let the dimensions of $R$ be $l \times h$, with the corresponding dimensions for $r$ being $\frac{l}{2} \times \frac{h}{2}$. Without loss of generality, we assume that $R$ and $r$ are axis-parallel. Using the results from Section \ref{sec:square-cover}, we generate a $1$-ply cover for $P$ using the inner rectangles (copies of $r$) in $O(n\log n)$ time. Finally, we replace the inner rectangles with the corresponding translations of $C$ in $O(mn)$ time. Thus, the overall process takes $O(n\log n + nm)$ time. This process results in a valid cover since $r \subseteq C$. Furthermore, since $C \subseteq R$, the ply value resulting from $C$ will be less than or equal to the ply value resulting after replacing the inner rectangles with the corresponding outer rectangles.

\begin{figure}[ht!]
    \centering
    \includegraphics[width=8cm]{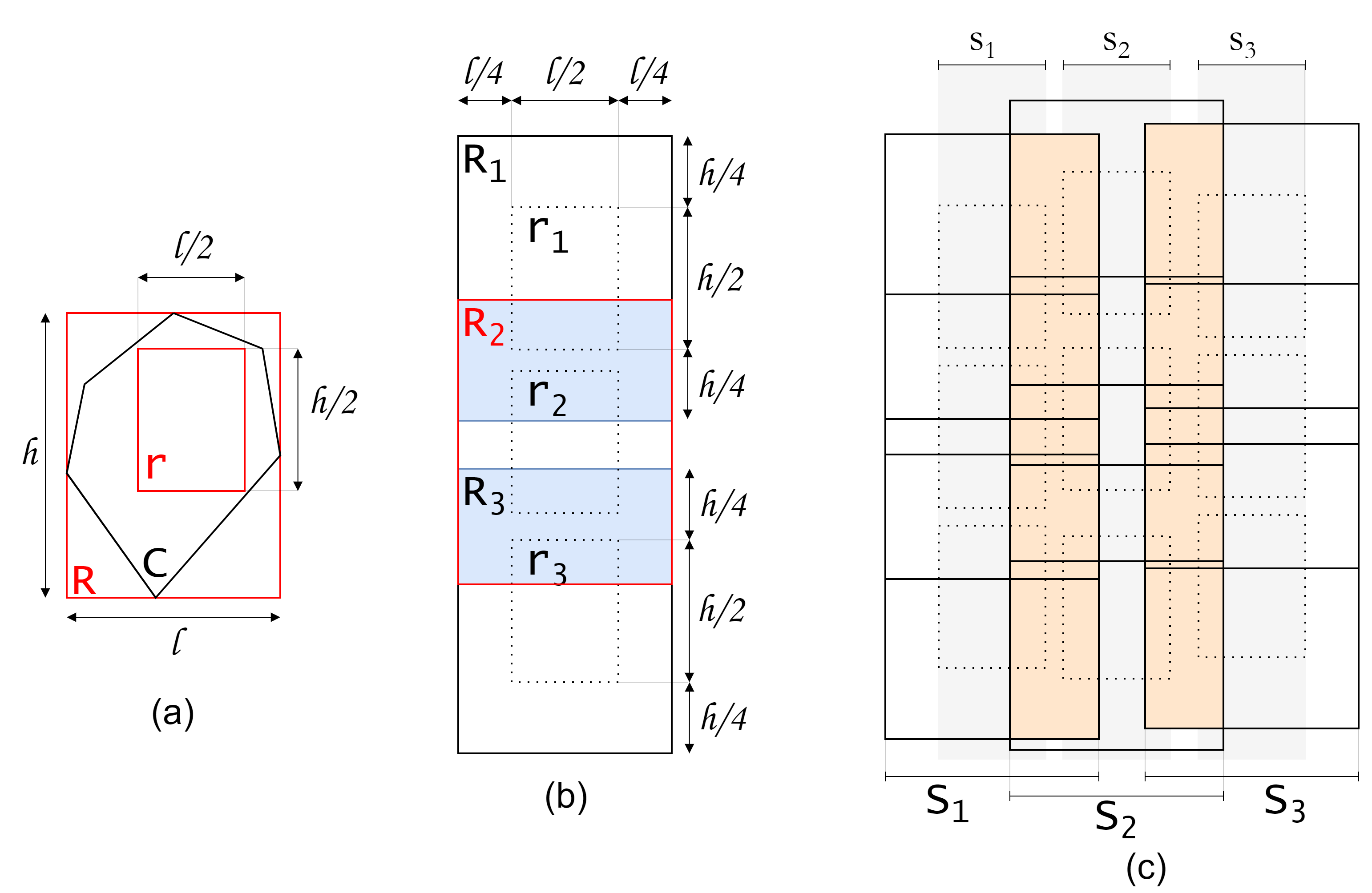}
    \caption{Illustrations for Convex Polygon Cover:
    (a) Approximating pair $(r, R)$ for polygon $C$. 
    (b) Configuration of a single vertical strip. 
    (c) Configuration of multiple vertical strips.
    }
    \label{fig:convex}
\end{figure}

To simplify the analysis, we assume that $r$ and $R$ are concentric, which can be achieved by equally shifting all outer rectangles such that they become concentric with their respective inner rectangles while keeping the overall structure of all outer rectangles identical, thus keeping the ply unchanged.

We start by analyzing the ply for each vertical strip generated by the rectangle cover algorithm. Since the inner rectangles form a $1$-ply cover, they are horizontally and vertically separated. As shown in Fig. \ref{fig:convex}(b), $r_1$ and $r_2$ are vertically separated, and $r_2$ and $r_3$ are also vertically separated. Since all $(r, R)$ pairs are concentric, $R_1$ and $R_3$ will be vertically separated. Hence, $R_1 \cap R_3 = \emptyset$, but there can be intersections between $R_1$ and $R_2$, i.e., only between adjacent rectangles (blue regions in Fig. \ref{fig:convex}(b)). Thus, for each vertical strip, the ply value is at most $2$.

Analogously, we can see that only adjacent vertical strips can intersect/touch. As shown in Fig. \ref{fig:convex}(c), $s_1, s_2$ and $s_3$ are vertical strips generated by the algorithm while computing the $1$-ply inner rectangle cover. $S_1, S_2$, and $S_3$ are vertical strips obtained by replacing inner rectangles with outer rectangles. Since $s_1, s_2$ and $s_3$ are disjoint, $S_1 \cap S_3 = \emptyset$. Hence, the maximum ply region will result from the intersection of adjacent vertical strips (orange regions in Fig. \ref{fig:convex}(c)). Since each vertical strip has a maximum ply value of $2$, the maximum ply value for the outer rectangle cover will be $4$. Hence, after replacing the inner rectangles with the corresponding translations of $C$, we get a ply value at most $4$. Hence, this is a $4$-ply convex polygon cover.
\end{proof}

\bibliographystyle{splncs04}
\bibliography{refs}

%Lemma \ref{lemma:hypercuber-cover} directly follows from claim \ref{claim:hypercube-cover} and claim \ref{claim:approx-hypercube}.

% \textbf{Referencing is working as expected above. Since the claims are not numbered in LNCS, therefore the claim number is not rendering.}

\end{document}